\documentclass[twocolumn,aps,showpacs,amssymb]{revtex4}

\usepackage{graphicx}
\usepackage{dcolumn}
\usepackage{bm}
\usepackage[colorlinks]{hyperref}

\begin{document}


\title{An approximative calculation of the fractal
structure in self-similar tilings}

\author{Yukio Hayashi}
\email{yhayashi@jaist.ac.jp}
\affiliation{
Japan Advanced Institute of Science and Technology,\\
Ishikawa, 923-1292, Japan
}


\begin{abstract}
Fractal structures emerge from statistical and 
hierarchical processes in 
urban development or network evolution. 
In a class of efficient and robust geographical networks, 
we derive the size distribution of layered areas, 
and estimate the fractal dimension by using the distribution 
without huge computations. 
This method can be applied to self-similar tilings based on 
a stochastic process.
\end{abstract}


\maketitle


\section{Introduction}
Fractal nature is observed in our real-life infrastructures of urban
spatial organization and many technological networks.
Indeed, the fractal dimensions of 
land-use and area-perimeter have been measured in real data of urban
cities \cite{Batty94}\cite{Frankhauser98}.
Similar structures with mixing of dense and sparse areas of nodes have
been found in router networks \cite{Yook02}, 
air transportation networks \cite{Guimera05}, 
and mobile communication networks \cite{Lambiotte08}.
These studies show that the spatial distribution of human activities is
inhomogeneous and concerning with the population density.
Thus, 
there commonly exists a hierarchical structure based on
statistical self-similarity beyond regular mathematical models such as 
Sierpinski gasket and carpet.
In other words, fractal behavior is not limited to objects with a
regular morphology, but can be introduced by iterative evolutionary
processes of subdivision or growing at different levels of scaling.

Recently, a multi-scale quartered (MSQ) network,
which is stochastically constructed by a self-similar tiling,
has been proposed \cite{Hayashi09}\cite{Hayashi10}.
The geographical network embedded on a planar space 
has several advantages \cite{Hayashi09}:
the robustness of connectivity against failures and attacks, 
the bounded short path lengths, 
and the decentralized routing algorithm \cite{Bose04}
 in a distributed manner.
Furthermore \cite{Hayashi10}, 
it is more efficient (economic) with shorter link lengths
and more suitable (tolerant) 
with lower load for avoiding traffic congestion 
than the state-of-the-art geometric growing networks
\cite{Zhang08,Wang06,Rozenfeld06,Dorogovtsev02,Zhou05,Zhang06,Doye05} 
and the spatially preferential attachment models 
with various topologies 
ranging from river to scale-free geographical networks.
These properties are useful for the future self-organized design of 
wide-area wireless ad hoc networks.

This paper investigates the size distribution of faces iteratively
divided in the generation process of a MSQ network.
In particular, we derive an approximative equation for the averaging
behavior of random process, and apply it to easily estimate the fractal
dimension of such a self-similar tiling.

\section{MSQ network} 
\begin{figure}[htb]
\includegraphics[height=37mm]{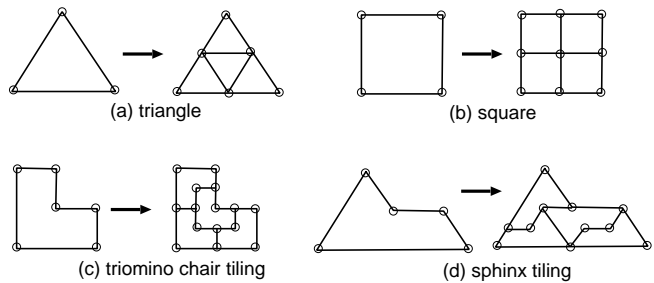}
\caption{Basic process of the division.}
\label{fig_basic_process}
\end{figure}

For generating a MSQ network, 
the following process is repeated from an initial tiling
which consists of same shaped faces.
At each time step, a face is chosen, e.g., with a probability 
proportional to the population in the space of a face
for the load balancing of communication requests, 
or uniformly at random. 
Then, as shown in Fig. \ref{fig_basic_process},
four smaller faces are created in the chosen face, and 
a planar network is self-organized on a geographical space.
Such a fractal structure 
is also observed in urban road networks \cite{Clauset06}
\cite{Cardillo06}. 
Note that 
the MSQ network includes a Sierpinski gasket obtained by
a special selection
when each triangle, except the central one, is hierarchically divided,
however its fractal dimension $\log 3 / \log 2 \approx 1.585$ differs to
that in the average behavior of the following random selection.

\section{Infinite state Markov chain} 
\begin{figure}[htb]
\includegraphics[height=60mm]{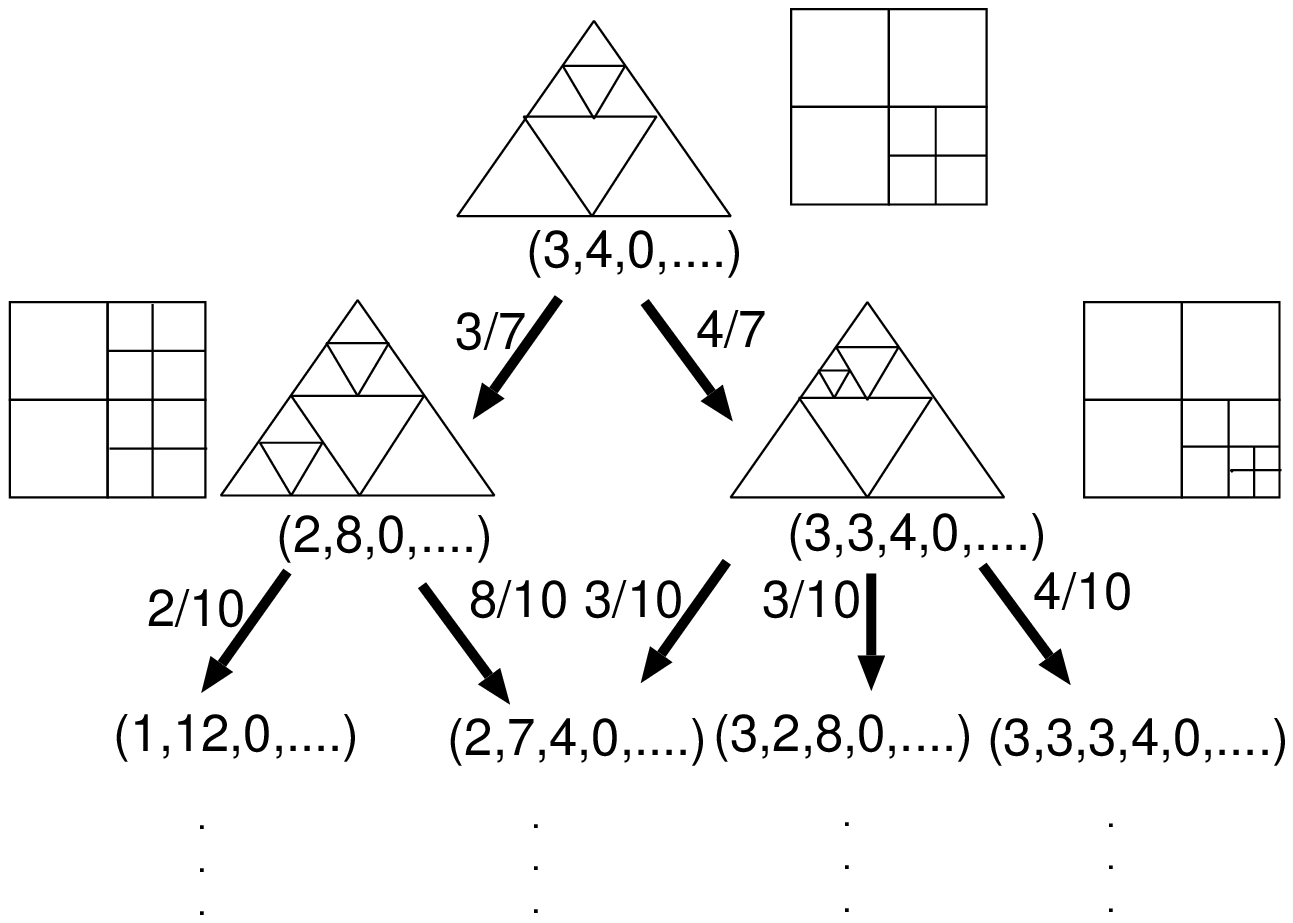}
\caption{Iterative division of a face chosen uniformly at random.}
\label{fig_process}
\end{figure}

For simplicity, 
we treat the uniformly random selection of a face, whose case 
corresponds to general positions of nodes
in the geographical network.
In addition, 
even if a face in the area of high population density
tend to be chosen for the subdivision, 
the assigned population to a node in its territory 
is asymptotically balanced,
then the behavior closes to the case of random selections 
in course of time.
In this setting, the subdivision process makes 
an infinite state Markov chain as illustrated in 
Fig. \ref{fig_process}. 
Here, we consider a vector 
$(n_{1}(t), n_{2}(t), \ldots, n_{l}(t) \ldots)$ 
as the state of Markov chain, 
whose element $n_{l}(t)$ represents the number of
faces at time step $t$
on the layer $l$ defined by decreasing order of size. 
A face on the layer $l$ is chosen with the probability
$p_{l}(t) = n_{l}(t) / {\cal N}(t)$ at the next step $t+1$, 
then the sate is changed from 
$(\ldots, n_{l}(t), n_{l+1}(t), \ldots)$ to 
$(\ldots, n_{l}(t+1)-1, n_{l+1}(t+1)+4, \ldots)$,
where ${\cal N}(t) = \sum_{k} n_{k}(t) = {\cal N}_{0} + 3 t$ 
denotes the total number of faces at $t$, 
and ${\cal N}_{0}$ is the initial number. 
Note that $p_{l}(t)$ is time-dependent on a selected path 
for the decision tree from the top to the bottom 
in Fig. \ref{fig_process},
equivalently on a selection sequence of faces.
Thus, the above Markov chain 
is different from the Galton-Watson type 
branching process with a time-independent 
probability for generating offsprings \cite{Liggett99}.

We obtain the averaging behavior for 
\begin{equation}
  \Delta n_{l} \stackrel{\rm def}{=} n_{l}(t+1) - n_{l}(t).
   \label{eq_delta_nl1} 
\end{equation}
This can be written as 
\begin{equation}
  \Delta n_{l} = 4 p_{l-1}(t) - p_{l}(t), \label{eq_delta_nl2} 
\end{equation}
since a face on the layer $l$ chosen with the probability $p_{l}(t)$ 
is divided into four smaller ones which belong to the layer $l+1$, 
therefore a face on the layer $l-1$ contributes 
to increase the number of faces on the layer $l$.
For a large $t$, by noticing $n_{l}(t) = {\cal N}(t) p_{l}(t)$ and 
substituting ${\cal N}(t) = {\cal N}_{0} + 3 t \approx 3 t$ 
into the right-hand side of Eq. (\ref{eq_delta_nl1}), it is 
\[
  \begin{array}{lll}
  \Delta n_{l} & = & 3(t+1) p_{l}(t+1) - 3t p_{l}(t), \\
               & = & 3t [ p_{l}(t+1) - p_{l}(t) ] + 3 p_{l}(t+1).
  \end{array}
\]
Using $p_{l}(t+1) \approx p_{l}(t)$ 
because of $t+1 \approx t \gg 1$, 
Eq. (\ref{eq_delta_nl2}) is rewritten to 
\begin{equation}
  p_{l}(t+1) - p_{l}(t) = - \frac{4}{3 t}[p_{l}(t) - p_{l-1}(t)].
\label{eq_delta_pl} 
\end{equation}

\begin{figure}[htb]
\includegraphics[height=85mm,angle=-90]{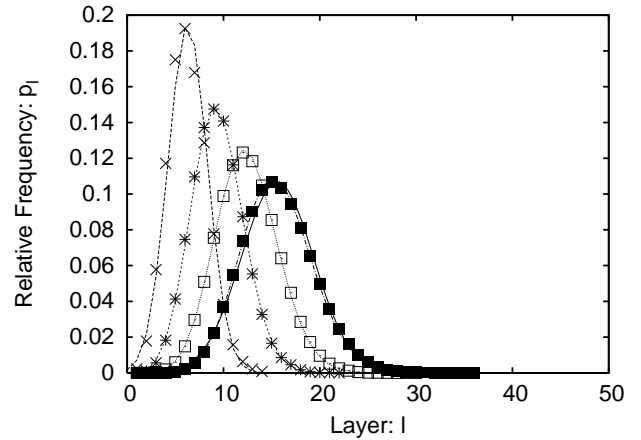}
\caption{The distribution of $p_{l}$.}
\label{fig_moving_wave_uniform}
\end{figure}

\begin{figure}[htb]
  \includegraphics[height=85mm,angle=-90]{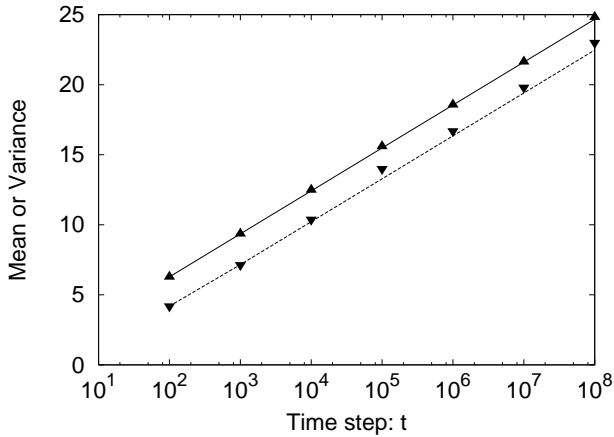}
  \caption{Mean \& Variance of the distribution of $p_{l}$.}
  \label{fig_mean_var}
\end{figure}

\begin{figure}[htb]
 \includegraphics[height=85mm,angle=-90]{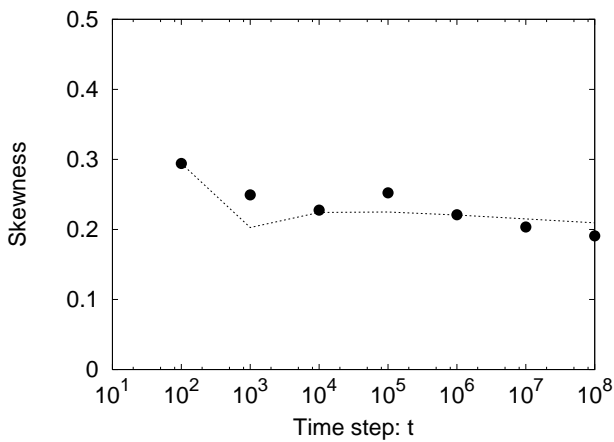}
 \caption{Skewness of the distribution of $p_{l}$.}
  \label{fig_skew}
\end{figure}

We numerically confirm 
that the average behavior 
of the Markov chain is well fitting to the solution for 
Eq. (\ref{eq_delta_pl}) as shown in Fig. \ref{fig_moving_wave_uniform}.
Here, the cross, asterisk, open square, and closed square marks
correspond to the average of 100 states reached 
at each time $t = 10^{2}, 10^{3}, 10^{4}$, and $10^{5}$, respectively, 
on the decision tree 
from the initial state $(3, 4, 0, \ldots)$ in Fig. \ref{fig_process}. 
The four types of dashed curves correspond to the distributions 
calculated by Eq. (\ref{eq_delta_pl}) until these times.
The solid curve fitting with the closed square marks 
is the approximation by a normal distribution.
Note that the averaged or calculated 
number $n_{l}$ of faces on each layer $l$ at $t$
is normalized as the relative frequency $p_{l}$.
This bell-shaped distribution means that  
there exist a mixing of dense and sparse areas with
various sizes of faces whose majorities have the
intermediate sizes. 
Figure \ref{fig_mean_var} shows 
the mean and the variance of the distribution $p_{l}$. 
The upper and lower triangle marks denote
   the mean and the variance of the distribution obtained
   for the Markov chain.
   The solid and dashed lines denote the corresponding results for 
   Eq. (\ref{eq_delta_pl}).
In Fig. \ref{fig_skew}, 
the circle marks and dotted piecewise linear line denote the skewness
   of these distributions for the Markov chain and for 
   Eq. (\ref{eq_delta_pl}), respectively.
The mean and the variance grow as $O(\log t)$, while the skewness is 
almost constant around 0.2. 
Although the $p_{l}$ is asymptotically a Poisson distribution with a
same value for the mean and the variance as
mentioned in the Appendix, 
it is rather fitting with the approximation 
by Eq. (\ref{eq_delta_pl}) in a finite size.

\begin{figure}[htb]
 \begin{center}
 \includegraphics[height=46mm]{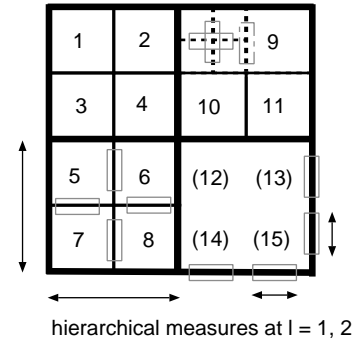}
 \caption{Illustration for a cover-based method.
 We count holes by the numbers for $n_{2}$ and the parenthesized numbers
  for $n_{1}$.} \label{fig_cover-based_squ}
 \end{center}
\end{figure}

On the basis of the numerically obtained $p_{l}(t)$, 
the fractal dimension $d_{f}$ is estimated as 
1.2 (a similar value is obtained 
for a chair or a sphinx tilling \cite{Solomyak97})
by using a cover-based method in the hierarchically coarse measure 
of doubling link length on each layer $l$ until $t = 10^{8}$.
Figure \ref{fig_cover-based_squ} shows an example at the state
$(1, 11, 4, 0, \ldots)$.
The number of covered areas (gray rectangles) is 
\[
  N[2] = 2 \times N[1] + 11 + 
  3 \times (4^{2} - (4 \times 1 + 11)), 
\]
where we set the number of initial coverings 
$N[1] = 12$ for a square tiling.
It depends on the primitive shape;
$N[1] = 9$ for a triangle tiling, and 
$N[1] = 8$ for a chair or a sphinx tiling.
In general for a large $l$, 
we calculate $d_{f} = \log N[l] / \log 2^{l}$ 
from 
\begin{equation}
   N[l] = 2 \times N[l-1] + n_{l} + 
   3 \times (T_{l} - S_{l}), 
   \label{eq_N_eps}
\end{equation}
the total number of faces in the completely recursive 
division  
$T_{l} \stackrel{\rm def}{=} 4^{l}$, 
the number of holes without subdivisions 
$S_{l} \stackrel{\rm def}{=} 4 \times S_{l-1} + n_{l}$, 
and $S_{1} = n_{1}$.
Three terms in the right-hand side of Eq. (\ref{eq_N_eps}) 
correspond to thick, thin solid lines, 
and dashed lines in Fig. \ref{fig_cover-based_squ}.
Since there is one-to-one correspondence between 
the divided four faces and the cross-edges in a square
at each layer $l$, 
e.g., by applying a clockwise mapping rule, 
$n_{l}$ represents this effect on the covering.
However, the 2nd and 3rd terms are replaced by 
$3 \times \lceil n_{l} / 4 \rceil$ for a triangle tiling, 
and by $8 \times \lceil n_{l} / 4 \rceil$ for 
a chair or a sphinx tiling.
We emphasis that our method 
can be performed by only using the calculated $n_{l}$ 
without both averaging 
and image processings in a box-counting method.

We should remark that 
the continuous approximation with respect to 
time and space variables for both
sides of Eq. (\ref{eq_delta_pl})
\[
  \frac{\partial p_{l}}{\partial t} = - \frac{4}{3 t} 
  \frac{\partial p_{l}}{\partial l}
\]
gives an incorrect solution, 
which keeps the shape in the traveling wave 
\[
  p_{l}(t) = F(l - v \log t), 
\]
where a function $F()$ is determined by the initial distribution, 
and $v = 4/3$.

\section{Conclusion} 
Considering the stochastic network construction 
\cite{Hayashi09}\cite{Hayashi10} as a Markov chain, 
the size distribution of the divided faces has been investigated.
We have shown, by numerically solving Eq. (\ref{eq_delta_pl})
as the average behavior,
that the mean and the variance are proportional to 
$\log t$, while the skewness is small positive.
In a realistic finite size ${\cal N} \propto t \leq 10^{8}$, 
this approximation of $p_{l}$ 
is fitting better than the asymptotical Poisson solution 
for the interactive particle system (see the Appendix).
From the distribution $n_{l} = {\cal N} p_{l} $, 
the fractal dimension is estimated as $1.2$ without huge computations of
averaging and image processings in the conventional 
box-counting method.
The proposed method can be applied to other self-similar tilings based
on a stochastic process.
Furthermore, it would be interesting to discuss a self-similar modeling
which reproduces the fractal structures of 
urban road \cite{Clauset06}\cite{Cardillo06}, 
air-sea transportation \cite{Guimera05}, 
communication \cite{Lambiotte08} networks.

\begin{acknowledgments}
The author would like to thank Prof. Norio Konno 
in Yokohama National University 
for suggesting the theoretical analysis in the Appendix.
This research is supported in part by a 
Grant-in-Aid for Scientific Research in Japan, No. 21500072.
\end{acknowledgments}

\appendix
\section{}
Let us reformulate the subdivision process for generating a MSQ network 
as a model in interacting infinite particle systems.
Then, according to the notation \cite{Liggett99}, 
we consider the generator $\Omega_{H}$ on 
$Z_{+} = \{ 1, 2, \ldots \}$,
\[
   \Omega_{H} f(\eta) = \sum_{x \in Z_{+}} \eta(x) 
   [ f(\eta^{x,x+1}) - f(\eta) ],
\]
where we set $m=4$ and define the following functions on 
$\eta \in \{ 0, 1, 2, \ldots\}^{Z_{+}}$, 
\[
   \eta^{x,x+1}(y) \stackrel{\rm def}{=} \left\{ 
   \begin{array}{ll}
     \eta(x+1) + m, & y = x+1,\\
     \eta(x) - 1,   & y=x,\\
     \eta(y),       & y \neq x, x+1.\\
   \end{array} \right.
\]
Note that the position $x$ corresponds to the layer $l$ 
of a chosen face in Sec. 3,
therefore the number $\eta(x)$ decreases by $1$ at $x$ and increases 
by $m$ at $x$ due to the contribution at $x-1$.
In particular for $f(\eta) = \eta(x)$, we have
\begin{equation}
  \frac{d n_{x}(\tau)}{d \tau} = m \times n_{x-1}(\tau) 
   - n_{x}(\tau), \;\; x \geq 1, \label{eq_IPS1}
\end{equation}
\begin{equation}
  \frac{d n_{0}(\tau)}{d \tau} = - n_{0}(\tau),  \label{eq_IPS2}
\end{equation}
where $n_{x}(\tau) = E[\eta_{\tau}(x)]$ 
is the expectation of the number of particles at position $x$ 
and at time $\tau$,
and the initial configuration is 
$\eta_{0} = (m, 0, 0, \ldots)$. 
We can solve (\ref{eq_IPS1})(\ref{eq_IPS2}),
\[
   n_{x}(\tau) = m^{x} \frac{\tau^{x-1}}{(x-1)!} e^{-\tau}, 
   \;\; x \geq 1.
\]
Also, the expectation of the total number of particles is given by
\[
   {\cal N}(\tau) = m e^{(m-1) \tau}.
\]
Therefore, from $p_{x}(\tau) = n_{x}(\tau)/ {\cal N}(\tau)$, 
we obtain the 
Poisson distribution with a parameter $m \tau$
\[
   p_{x}(\tau) = \frac{(m \tau)^{x-1}}{(x-1)!} e^{- m \tau}, 
   \;\; x \geq 1.
\]
By the variable transformation between $t$ and $\tau$ 
from the relation 
${\cal N}(t) = {\cal N}(0) + (m-1)t 
\Leftrightarrow {\cal N}(\tau) = m e^{(m-1) \tau}$,
the mean and the variance follow $m \tau \propto \log t$. 
This logarithmic behavior 
is consistent with the numerical results in Sec. 3.

As another typical approach, if we consider a generating function 
\[
  H^{(t)}(z) \stackrel{\rm def}{=} \sum_{l = 1}^{\infty} p_{l}(t) z^{l},
\]
by multiplying $z^{l}$ to both sides of Eq. (\ref{eq_delta_pl})
and taking the summation $\sum_{l = 1}^{\infty}$, 
we obtain 
\[
  H^{(t+1)}(z) = \left( 1 - \frac{4 (1 - z)}{3 t}
                 \right) H^{(t)}(z).
\]
Here, we use $\sum_{l} p_{l-1}(t) z^{l} = z  H^{(t)}(z)$.
The frequency $ p_{l}(t)$ is only formally 
calculated by the recursive differentials
\[
  p_{l}(t) = \left. \frac{1}{l !} 
                 \frac{\partial^{l} H^{(t)}(z)}{\partial z^{l}} 
               \right|_{z = 0}.
\]
However, it is practically unsolvable 
because of involving with a very complicated 
combinatorial explosion for a large $t$, 
equivalently for a large network size $N \propto t$.

\end{document}